\documentclass[11pt,a4 paper]{article}
\usepackage{cite}
\usepackage{amssymb}
\usepackage{amsmath}
\usepackage[dvips]{graphicx}
\begin{document}
\thispagestyle{empty}
\begin{center}
\noindent {\textbf{\Large Holographic Dark Energy Model in Brans-Dicke Theory with Future Event Horizon}}
\end{center}
\vspace{1cm}
\begin{center}

\noindent  \textbf{\textbf{Milan Srivastava\;$^{2}$ \;and \; C. P. Singh\;\footnote{Corresponding author}}}\\

\vspace{0.5cm}

\noindent{ $^{1,2}$Department of Applied Mathematics,\\
 Delhi Technological University \\
 Bawana Road, Delhi-110 042, India.}\\
\texttt{ $^1$cpsphd@rediffmail.com  \\
$^2$milandtu@gmail.com}\\
\end{center}
\vspace{1.5cm}
\noindent {\textbf{Abstract}} In this paper, we study the dynamics of non-interacting and interacting holographic dark energy models in the framework of Brans-Dicke theory. As system's infra-red cut-off we consider the future event horizon. The motivation of this work is to use the logarithmic form of the Brans-Dicke scalar field, $\phi \propto ln(\alpha+\beta a)$, where $\alpha$ and $\beta$ are constants and `a' is the scalar factor  as proposed in a recent work \cite{kumar} to study the new agegraphic dark energy models. We find the time-dependent equation of state parameter and deceleration parameter which describe the phase transition of the universe. We observe that the model explains the early time inflation and late time acceleration including matter-dominated phase. It is also observed that the equation of state parameter may cross phantom divide line in late time evolution. The cosmic coincidence problem is also discussed for both the models. We observe that this logarithmic form of Brans-Dicke scalar field is more appropriate to achieve a less acute coincidence problem in non-interacting model whereas a soft coincidence can be achieved if coupling parameter in interacting model has small value.\\

\noindent \textbf{PACS NO: } 04.20.-q; 04.20.-Jb.\\
\noindent \textbf{Keywords-} Cosmology; FRW model; Holographic dark energy; Brans-Dicke theory;  Coincidence problem.
\pagebreak

\pagestyle{myheadings}
\section{Introduction}
\indent The astrophysical observations \cite{riess,perl,ben,kom,per} have led to many interesting phenomena in the field of cosmology, especially, the accelerating expansion of the Universe. This event has brought many challenges to the cosmologists. The observations show that the accelerating expansion of the Universe is due to a mysterious form of energy with negative pressure which is dubbed as ``dark energy" (DE). Nowadays, the DE has become one of the most active field in physics and astronomy. Many models have been proposed to describe this late -time acceleration of the Universe. The simplest candidate for DE is the cosmological constant\cite{car}. However, it suffers the so-called cosmological constant (CC) problem (the fine-tuning problem) and the cosmic coincidence problem \cite{wein,copeland}. \\
\indent In order to alleviate the CC problems and explain the accelerated expansion, some alternative models have been proposed either by modifying the right hand side of the Einstein field equations by considering the specific forms of energy -momentum tensor $T_{\mu\nu}$ or modifying the left hand side of Einstein field equations. The DE models which belong to the first category include quintessence \cite{caro}, k-esssence \cite{chiba}, the family of chaplygin gas \cite{kam}, holographic \cite{hooft,suss}, new agegraphic \cite{wei}, etc. The models which belong to second category are the modified gravity that include $f(R)$ gravity \cite{copoz}, scalar -tensor theories \cite{amen}. The poineering study on scalar-tensor theories was done by Brans and Dicke \cite{brans} to incorporate Mach's principle into gravity which is known as Brans-Dicke (BD) theory. This was the first gravity theory in which the dynamics of gravity were described by a scalar field while spacetime dynamics were represented by the metric tensor. In BD theory, the gravitational constant $G$ is replaced with the inverse of a time-dependent scalar field $\phi$, namely, $\phi=(8\pi G)^{-1}$, and this scalar field couples to gravity with a coupling parameter $\omega$. Since the BD theory passes the experimental tests from the solar system \cite{bert} and provides an explanation of the accelerated expansion of the Universe, it is worthwhile to discuss DE models in this framework.\\
\indent In recent years, the holographic dark energy (HDE) \cite{li,hsu,setare1,setare2,ncruz,jzhang,sheykhi} has been studied as a possible candidate for dark energy. This type of model was motivated by the holographic principle \cite{hooft} which might lead to the quantum gravity to explain the events involving high energy scale. According to holographic principle, the number of degrees of freedom in a bounded system should be finite and has relations with the area of its boundary. For a system with size $L$ and ultra-voilet (UV) cut-off $\Lambda$ without decaying into a black hole, it is required that the total energy in a region of size $L$ should not exceed the mass of a black hole of the same size, thus $L^3\rho_{\Lambda}\le L M_P^2$. The largest $L$ allowed is the one saturating this inequality, thus the holographic DE density $\rho_{\Lambda} = 3c^2 M_P^2 L^{-2}$ where $c$ is a numerical constant and $M_P$ is the reduced Planck mass $M_P^{2}=1/8\pi G$. It means that there must be a duality between UV cut-off and (infra-red) IR cut-off. Therefore, the UV cut-off is related to the vacuum energy and IR cut-off is related to the large scale of the Universe. The large scale of the Universe can be taken as, for example Hubble horizon, particle horizon or event horizon \cite{li,hsu}.  Nowadays, the HDE model is the most efficient DE candidate among all the existing DE candidates which is also capable to explain the acceleration of the Universe.  It has been shown that the HDE model is favored by the latest observational data including the sample of Type Ia supernovae (SN Ia), the shift parameter of the cosmic microwave background (CMB), and the baryon acoustic oscillation (BAO) measurement \cite{qwu}. \\
\indent Li \cite{li} has discussed three choices for the length scale L which is supposed to provide an IR cutoff. The first choice is to identify $L$ with the Hubble radius, $H^{-1}$. In the formalism of HDE, the Hubble horizon is a most natural choice for the IR cut-off, but Hsu \cite{hsu} in GR and Xu et al. \cite{xu} in BD theory have shown that the Hubble horizon as an IR cutoff is not a suitable candidate to explain the recent accelerated expansion. The second option is the particle horizon radius but it also does not give an accelerated expansion. The third choice is the identification of $L$ with the radius of the future event horizon which gives the desired result, namely a sufficiently negative equation of state to obtain an accelerated universe.\\
\indent In this paper, we consider the HDE model with IR cut-off as future event horizon in BD theory. Since most of the authors \cite{setare1,sheykhi,ban} have discussed HDE model in BD theory with power-law form of the BD scalar field. In a recent paper, Kumar and Singh \cite{kumar} have revisited these papers and have proposed a logarithmic form of BD scalar field to discuss new agegraphic dark energy model in BD theory. The authors  have also claimed to resolve the cosmic coincidence problem. Actually, it has been noticed in  papers \cite{setare1,sheykhi,ban} that the  power-law form of BD scalar field gives the constant value of deceleration parameter whatever may be matter content. Therefore, power-law assumption can not describe the phase transition of the Universe. \\
\indent In this work, we revisit the study of Ref.\cite{sheykhi} and consider HDE model in BD theory with logarithmic form of BD scalar field as proposed in Ref. \cite{kumar}. We obtain the time -dependent deceleration parameter and equation of state parameter which describe the phase transition of the evolution of the Universe. We further discuss a cosmological model where the pressureless dark matter and HDE do not conserved separately but interact with each other. We also discuss the cosmic coincidence problem which has not been discussed in previous work \cite{setare1,sheykhi}. It is found that the HDE model successfully resolves the cosmic coincidence problem in BD theory with event horizon.\\
\indent The paper is organized as follows: In Sec 2, we present the field equations of HDE model in BD theory. We study the non-interacting holographic dark energy with IR cut-off as future event horizon. We calculate the equation of state parameter and deceleration parameter to discuss the nature of the evolution of the Universe. Section 3 deals with the interacting HDE model in BD theory. Section 4 gives the conclusion of the work.
\section{Holographic dark energy in Brans-Dicke Theory}
\noindent The modified Einstein-Hilbert (EH) action for the BD theory is given by \cite{brans}
\begin{equation}
S=\int d^{4}x\sqrt{g}\left(-\varphi R+\frac{\omega}{\varphi }g^{\mu\nu}\partial_{\mu}\varphi  \; \partial_{\nu}\varphi +\mathcal{L}_{m}\right),
\end{equation}
where $R$ denotes the Ricci scalar curvature, $\omega$ is a dimensionless coupling parameter between scalar field and gravity called BD parameter and $\mathcal{L}_{m}$ represents the matter Lagrangian density. On re-defining the scalar field $\varphi$ as
\begin{equation}
\varphi = \frac{\phi^2}{8 \omega}
\end{equation}
The action (1) of BD theory in the canonical form is given by \cite{setare1,arik, cali}
\begin{equation}
S = \int d^4 x \sqrt{g} \left( -\frac{1}{8 \omega}\phi^2 R + \frac{1}{2}g^{\mu\nu}\partial_{\mu}\phi  \; \partial_{\nu}\phi + \mathcal{L}_{m} \right),
\end{equation}
where $\phi$ is a time-dependent scalar field called BD scalar field. The non-minimal coupling term $\phi^2 R$ replaces with the EH term $R/G$ in such a way that $G^{-1}_{eff}=2\pi \phi^2/\omega$, where $G_{eff}$ is the effective gravitational constant as long as the dynamical scalar field $\phi$ varies slowly.\\
\indent We consider a homogeneous and isotropic Friedmann-Robertson-Walker (FRW) Universe, which is described by the line element
\begin{equation}
ds^{2}=dt^{2}-a^{2}(t)\left[ \frac{dr^2}{1-kr^2}+r^2(d\theta^2+\sin^2\theta d\Phi^2)\right],
\end{equation}
where $a(t)$ is the cosmic scale factor of the Universe and $k$ is the curvature parameter with $k = -1, 0, 1$ corresponding to open, flat and closed Universes, respectively. We assume that the Universe is filled with perfect fluid containing pressureless dark matter (DM) (excluding baryonic matter) and HDE.\\
\indent Taking the variation of action (3) with respect to the metric tensor, $g_{\mu\nu}$ for the line element (4), one can obtain the field equations for the non-flat universe containing HDE and DM as
\begin{equation}
\frac{3}{4\omega}\left( H^2+\frac{k}{a^2}\right) -\frac{1}{2}\frac{\dot\phi^2}{\phi^2}+\frac{3H}{2\omega} \frac{\dot\phi}{\phi} = \frac{\rho_m+\rho_h}{\phi^2},
\end{equation}
\begin{equation}
\frac{1}{4\omega}\left(2\frac{\ddot a}{a}+H^2+\frac{k}{a^2} \right)+\frac{H}{\omega}\frac{\dot\phi}{\phi}+\frac{1}{2\omega}\frac{\ddot \phi}{\phi}+\frac{1}{2}\left(1+\frac{1}{\omega}\right)\frac{\dot\phi^2}{\phi^2}=-\frac{p_h}{\phi^2},
\end{equation}
where $H=\frac{\dot a}{a}$ is the Hubble parameter, $\rho_{m}$ is the DM energy density, $\rho_{h}$ is the HDE energy density and $p_{h}$ is the pressure of HDE. Here, an over dot denotes the derivative with respect to the cosmic time $t$. The wave equation for the scalar field $\phi$ is given by
\begin{equation}
\ddot\phi+3H\dot\phi-\frac{3}{2\omega} \left( \frac{\ddot a}{a}+H^2+\frac{k}{a^2}\right)\phi = 0.
\end{equation}
In frame work of BD theory, the HDE density has the form
\begin{equation}
\rho_h=3 c^2 \varphi L^{-2},
\end{equation}
where $\varphi= M_P^2 = (8\pi G_{eff})^{-1}$. In canonical form $G_{eff}^{-1}= \frac{2\pi}{\omega} \phi^2$, the HDE density (8) now becomes
\begin{equation}
\rho_h=\frac{3c^2\phi^2}{4\omega L^2}.
\end{equation}
\indent It is to be noted that if the IR cut-off is taken as the Hubble horizon then the energy density of HDE and the critical density match identically. This situation generally arises in inflation scenario where $L=H^{-1}$. There are various other choices of IR cut-off for the cosmological length scale $L$ available in the literature, such as particle horizon, event horizon, Ricci length, Granda-Oliveros cut-off etc. However, particle horizon is not suitable to derive the acceleration. A suitable choice of future event horizon as an IR cut-off was suggested by Li et al \cite{M}. The cosmological length $L$ for the event horizon is defined as
\begin{equation}
L=ar(t),
\end{equation}
where the function $r(t)$ can be obtained from the relation
\begin{equation}
\int_0^{r(t)} \frac{dr}{\sqrt{1-kr^2}}=\int_t^{\infty}\frac{dt}{a(t)} = \frac{R_E}{a(t)}.
\end{equation}
where $R_E$ is the event horizon, defined by
\begin{equation}
R_E=a\int_t^{\infty}\frac{dt}{a(t)}=a\int_a^{\infty}\frac{da}{Ha^2}.
\end{equation}
The general solution of $r(t)$ from Eq. (11) for non-flat FRW model is given by
\begin{equation}
r(t)=\frac{1}{\sqrt{k}} \sin y,
\end{equation}
where, $y=\frac{\sqrt{k} R_E}{a(t)}$.\\
The critical energy density $\rho_{cr}$ and the energy density of the curvature $\rho_k$ are, respectively, defined as
\begin{equation}
\rho_{cr} = \frac{3 \phi^2 H^2}{4 \omega},
\end{equation}
\begin{equation}
\rho_{k} = \frac{3 k \phi^2}{4 \omega a^2}.
\end{equation}
It is useful to express the equations and physical quantities in the terms of fractional energy densities to analyse the results in better way. The fractional energy densities in their usual form are given as
\begin{equation}
\Omega_m = \frac{\rho_m}{\rho_{cr}} = \frac{4\omega \rho_m}{3\phi^2 H^2},
\end{equation}
\begin{equation}
\Omega_k = \frac{\rho_k}{\rho_{cr}} = \frac{k}{H^2 a^2},
\end{equation}
\begin{equation}
\Omega_h = \frac{\rho_h}{\rho_{cr}} = \frac{c^2}{H^2 L^2}.
\end{equation}
Equation (18) can also be written as
\begin{equation}
HL=\frac{c}{\sqrt{\Omega_h}}
\end{equation}
Now, differentiating (10) with respect to the cosmic time $t$, and using (13) and (19), we obtain
\begin{equation}
\dot L = HL + a \dot{r(t)} = \frac{c}{\sqrt{\Omega_h}} - \cos y.
\end{equation}
Let us first consider the case where HDE and DM do not interact. In this case, the conservation equations for HDE and DM are respectively given by
\begin{equation}
\dot \rho_h + 3 (1+w_h) \rho_h H = 0,
\end{equation}
\begin{equation}
\dot \rho_m + 3 \rho_m H = 0,
\end{equation}
where $w_h=\frac{p_h}{\rho_h}$ is the equation of state (EoS) parameter of HDE.\\
\noindent In order to discuss the physical behaviors of HDE model, we consider the following well-motivated ansatz for BD scalar field $\phi$ \;\cite{kumar}
\begin{equation}
\phi = \phi_0 ln(\alpha+\beta a),
\end{equation}
\noindent where $\phi_0>0$, $\alpha>1$ and $\beta>0$ are constants.\\
\indent This logarithmic form of $\phi$ fulfills the requirement of slow variation of $G$ as $\phi$ evolves slowly. It is important to notice that BD cosmology becomes standard cosmology when $\beta\rightarrow 0$. As we know that the universe exhibits phase transition from the past decelerated phase to the current accelerated phase. A time-dependent deceleration parameter (DP) is required to describe this phase transition. Therefore, the choice (23) of the  logarithmic form of BD scalar field, which always gives time-dependent DP, is physically acceptable. Kumar and Singh \cite{kumar} have discussed interacting new agegraphic dark energy (NADE) model with the Hubble horizon as an IR cut-off in the frame work of BD theory by assuming the above form of BD scalar field.\\
\indent In this paper, we discuss the HDE cosmological model with future event horizon as an IR cut-off in the frame work of BD theory as Li et al \cite{M} suggested that this IR cut-off is suitable to describe the accelerated expansion of the universe. \\
\indent On taking the derivative of (23) with respect to time, we get
\begin{equation}
\dot \phi = \phi_0 \frac{\beta a H}{(\alpha + \beta a)},
\end{equation}
\begin{equation}
\ddot \phi = \phi_0\left\{\frac{\beta a \dot H}{(\alpha + \beta a)}+\frac{\beta a H^2}{(\alpha+\beta a)}-\frac{\beta^2 a^2 H^2}{(\alpha + \beta a)^2}\right\}.
\end{equation}
\noindent By use of (20) and (24), Eq.(9) gives
\begin{equation}
\dot \rho_h = 2 H \rho_h\left(-1+\frac{\sqrt{\Omega_h}}{c} \cos y + \frac{\beta a}{(\alpha+\beta a) \;ln(\alpha+\beta a)}\right).
\end{equation}
\noindent Using (26) into (21), we obtain the EoS parameter for HDE
\begin{equation}
w_h = -\frac{1}{3} -\frac{2\beta a}{3(\alpha+\beta a) \; ln(\alpha+\beta a)}-\frac{2\sqrt{\Omega_h}}{3c} \cos y.
\end{equation}
\noindent The EoS $w_h$ is bounded from below by
\begin{equation}
w_h = -\frac{1}{3} -\frac{2\beta a}{3(\alpha+\beta a) \; ln(\alpha+\beta a)}-\frac{2\sqrt{\Omega_h}}{3c}.
\end{equation}
From (27), we observe that for $\beta=0$ the EoS parameter $w_h$ of HDE reduces to its respective form of non-flat standard cosmology \cite{huang}, which is given by
\begin{equation}
w_h = -\frac{1}{3} -\frac{2\sqrt{\Omega_h}}{3c} \cos y.
\end{equation}
\indent It is to be noted that for power-law form $\phi\propto a^{m}$ of BD scalar field \cite{sheykhi,setare1}, the second term in the value of $w_h$ of Eq.(28) is a constant term $\frac{2m}{3}$ whereas we get a time-dependent term $\frac{2\beta a}{3(\alpha+\beta a) \; ln(\alpha+\beta a)}$ due to logarithmic form of BD scalar field. Therefore, the value of $w_h$ in our model is more dynamic in comparison to power-law in BD theory.\\
\indent Let us discuss the behavior of EoS parameter $w_h$ of HDE as obtained in Eq.(28). It is clear that $w_h$ always has a negative value such that $w_h<-\frac{1}{3}$. The value of $w_h$ at the beginning of the evolution, i.e., $a=0$ is same as in Eq. (29) for GR because the second term is zero at $a=0$. Thus, we find a negative value of $w_h$ in the very early Universe. It is observed that the second term attains to maximum value during the evolution and approaches to zero in late time evolution. The maximum value depends on sufficiently small values of $\alpha$ and large values of $\beta$ during the process of evolution. However, the maximum value only depends on the parameter $\alpha$ and it is found that max$\left\{ \frac{\beta a}{(\alpha+\beta a) \; ln(\alpha+\beta a)}\right\}\rightarrow 1$ as $\alpha\rightarrow 1$, i.e., $w_h < -1$. Therefore, $w_h$ may cross phantom divide line for this condition. It may also possible that $w_h$ crosses the phantom divide line for $\sqrt{\Omega_h}>c[1-\frac{\beta a}{(\alpha+\beta a)\; ln(\alpha+\beta a)}]$, i.e., for both conditions HDE model may cross the phantom divide line ($w_h=-1$) and approaches to the phantom region.\\
\indent It is also interesting to note that as the logarithmic term converges to zero in the late-time of evolution and also $\cos y \rightarrow 1$ in late-time as $a\rightarrow \infty$, the EoS parameter starts behaving like its respective form in standard general relativity \cite{huang} and it will depend only on the values of $\Omega_h$ and $c$. The form of $w_h$ in the late-time is given by
\begin{equation}
w_h = -\frac{1}{3}-\frac{2 \sqrt{\Omega_h}}{3 c}.
\end{equation}
\indent In this model, the value of parameter $c$ determines the property of HDE in late time. Since, the observation predicts $\Omega_h \rightarrow 1$ for the present time, therefore, at $c=1$, $w_h$ approaches to $-1$, i.e., our model behaves like cosmological constant. We get $w_h>-1$ but less than $-1/3$ at $c>1$, i.e., our model shows the quintessence region and if $c<1$,  we get $w_h<-1$, i.e., the phantom type behaviour occur. Thus, we conclude that when $c>1$, $c=1$ and $c<1$, one can generate quintessence, cosmological constant and  phantom respectively for non-interacting HDE model in BD theory.\\
\indent Now, we study the behaviour of deceleration parameter (DP) to discuss the evolution of the Universe. The DP, $q=-\frac{a\ddot{a}}{\dot{a}^2}=-1-\frac{\dot H}{H^2}$, can be obtained after dividing Eq. (6) by $H^2$, and using Eqs. (9), (17), (19), (24) and (25), which is given as
\begin{equation}
q=\frac{1+\Omega_k+3 w_h \Omega_h +\frac{4\beta a}{(\alpha+\beta a) \;ln(\alpha+\beta a)}-\frac{2\beta^2 a^2}{(\alpha+\beta a)^2 \;ln(\alpha+\beta a)}+\frac{2(\omega +1)\beta^2 a^2}{(\alpha+\beta a)^2 \; [ln(\alpha+\beta a)]^2}}{2\left(1+\frac{\beta a}{(\alpha+\beta a)\; ln(\alpha+\beta a)}\right)}.
\end{equation}
\indent The term $\frac{\beta^2 a^2}{(\alpha+\beta a)^2 \; ln(\alpha+\beta a)}$ has the same behaviour as the term $\frac{\beta a}{(\alpha+\beta a) \; ln(\alpha+\beta a)}$ except it has maximum value lies in (0, 0.41) depending on the value of $\alpha$. The BD parameter $\omega$ also plays an important role in the value of $q$. The solar system experiment Cassini gave a very high bound on $\omega$ as $|\omega|>40000$ \cite{berto,will}, whereas the cosmological observations provide the relatively lower bounds on $\omega$ \cite{Xu,aqua,wu,Li}. The observations suggest that $\omega$ has the large value so the last term of numerator of Eq. (31) containing $\omega$ will dominate during the evolution of the Universe. This shows $q$ may attain some positive value, i.e., the decelerated expansion of the Universe may occur during the evolution. Thus, HDE model explains the matter dominated phase of the Universe. As in the late-time of evolution, $w_h\rightarrow -1$ and the terms $\frac{\beta a}{(\alpha+\beta a) \; ln(\alpha+\beta a)}$ and $\frac{\beta^2 a^2}{(\alpha+\beta a)^2 \; ln(\alpha+\beta a)}$ converge to zero as $a\rightarrow \infty$. Then, the value of $q$ in the late time of evolution is obtained as
\begin{equation}
q\approx\frac{1+\Omega_k -3 \Omega_h}{2}.
\end{equation}
\indent Since the observations recommend that our present Universe is almost flat, i.e., $k=0$ ($\Omega_k=0$), Eq.(32) gives $q\approx \frac{1-3 \Omega_h}{2}$. Thus, it is observed that  $q$ is negative for $\Omega_h>1/3$, i.e., the accelerated expansion is obtained for $\Omega_h>1/3$. It can also be noticed that if we consider the open Universe, i.e., $k<0$ ($\Omega_k<0$), the accelerated expansion can be obtained more easily. Even for the closed geometry case of the Universe we can also get an accelerated expansion of the Universe but for this we must have a very large value of $\Omega_h$ which will give a negative value of $q$. Thus, we can conclude that the HDE model describes the phase transition from early time inflation to the matter dominated phase and then matter dominated phase to late time accelerated phase.\\
\indent Let us consider the cosmological coincidence problem which was raised first time by Steinhardt \cite{stein,zla}. The problem may be resolved by making the density ratio $r_1=\frac{\rho_m}{\rho_h}$ is of order unity, i.e., $(r_1)_0 \sim \mathcal{O}(1)$ for a wide range of initial condition. The second way is that either $r_1$ converses to a constant value or evolve very slowly in late-time of evolution. From (5), the energy density ratio is given by
\begin{equation}
r_1=-1+\frac{1}{\Omega_h}\left[ \Omega_k +1-\frac{2\;\omega \beta^2 a^2}{3(\alpha+\beta a)^2 \; [ln(\alpha+\beta a)]^2} +\frac{2\beta a}{(\alpha+\beta a) \;ln(\alpha+\beta a)} \right].
\end{equation}
\indent Therefore, we obtain a time-dependent value of $r_1$. At the beginning  of evolution the value of $r_1$ is $\left\{-1+\frac{(\Omega_k +1)}{\Omega_h}\right\}$ as the last two terms vanish at $a=0$. In the late time of evolution we obtain the same expression of $r_1$ as the last two terms approaches to zero as $a\rightarrow \infty$. Now, the evolution of energy density ratio $r_1$ can be obtained as
\begin{equation}
\dot r_1= 3 r_1 H w_h.
\end{equation}
\indent According to $\Lambda$CDM model, $r_1$ evolves as $|\frac{\dot r_1}{r_1}|_0 = 3H_0$. Throughout the paper the subscript zero represents the present value of the quantity. Since, for $c=1$ our model shows $w_{h0} = -1$ in late-time, therefore, we get $|\frac{\dot r_1}{r_1}|_0 = 3H_0$ which is same as for $\Lambda$CDM model. This shows that there is no reduction in the acuteness of the coincidence problem. Since, the EoS parameter $w_h$ is time-dependent value, therefore, the less acute coincidence problem can be obtained if we have a quintessence like EoS parameter ($w_h>-1$). Also, we may achieved $w_{h0} > -1$ for $\sqrt{\Omega_h}<\frac{c}{\cos y}[1-\frac{\beta a}{(\alpha+\beta a)\; ln(\alpha+\beta a)}]$. Since, the second term converges to zero as $a\rightarrow \infty$, we can get quintessence like EoS parameter more conveniently due to the logarithmic form as compare to power law form of BD scalar field whereas we get the constant second term. Thus, we can conclude that this logarithmic form of BD scalar field is more appropriate to achieve a less acute coincidence problem. Now, let us assume $w_{h0}=-2/3$ we obtained $|\frac{\dot r_1}{r_1}|_0 = 2H_0$. Clearly, it shows less acuteness in the coincidence problem as compare to the $\Lambda$CDM model. But, the problem is more acute in the case of phantom like EoS parameter ($w_{h0} < -1$). This case shows the more complex condition of coincidence problem as compare to the $\Lambda$CDM model.\\
\section{Interacting HDE in Brans-Dicke Cosmology}
\indent In this section, we extend our study to the case where both dark components, the pressureless DM and the HDE, interact with each other. If we proceed to consider a scenario of interacting dark energy, $\rho_m$ and $\rho_h$ do not satisfy independent conservation laws, they instead satisfy
\begin{equation}
\dot \rho_h+3 H (1+w_h)\rho_h = -\Gamma,
\end{equation}
\noindent and
\begin{equation}
\dot\rho_m +3 H\rho_m = \Gamma,
\end{equation}
\noindent where $\Gamma=3 b^2 H (\rho_m+\rho_h)$ is a particular interacting term \cite{karwan} with the coupling constant $b^2$. This interacting term can be rewritten in terms of ratio of density parameter $r_1=\rho_m/\rho_h$ as
\begin{equation}
\Gamma=3 b^2 H \rho_h(1+r_1),
\end{equation}
Using (33) and (37) into (35), the EoS parameter of HDE is given by
\begin{eqnarray}
w_h &=& -\frac{1}{3}-\frac{2\sqrt{\Omega_h}}{3c} \cos y-\frac{2\beta a}{3(\alpha+\beta a)\; ln(\alpha+\beta a)} \notag\\
& - &\frac{b^2}{\Omega_h}\left[1+\Omega_k+\frac{2\beta a}{(\alpha+\beta a)\; ln(\alpha+\beta a)}-\frac{2\omega\beta^2 a^2}{3(\alpha+\beta a)^2\;[ln(\alpha+\beta a)]^2}\right],
\end{eqnarray}
\noindent which is a time-dependent value. It is to be noted that the term $\frac{\beta a}{(\alpha+\beta a)\; ln(\alpha+\beta a)}$ has the same behaviour as discussed earlier for non-interacting case.  It is easy to find out that, in the limit of $\beta \rightarrow 0$, the standard cosmology is recovered. In the beginning of the evolution, the term $\frac{\beta a}{(\alpha+\beta a)\; ln(\alpha+\beta a)}$ is zero and  hence the EoS parameter of HDE (38) gives
\begin{equation}
w_h=-\frac{1}{3}-\frac{2\sqrt{\Omega_h}}{3 c} \cos y -\frac{b^2(1+\Omega_k)}{\Omega_h},
\end{equation}
which is same as the standard non-flat HDE model in BD theory. We find that $w_h$ is always  negative and less than $-1/3$ in the early of the evolution which shows the inflation in early time. The solar system experiments \cite{berto} predict a very high bound value of $\omega$ which is $|\omega|> 40000$. However, Acquaviva and Verde \cite{acq} found that $\omega$ may be smaller than 40000 in cosmological scale. Due to a large value $\omega$ suggested by the experiments, the last term containing $\omega$ will dominate during early phase of the evolution of the universe. Thus, we observe a positive value of $\omega$. Thus, the decelerated phase occurs during the evolution of the universe. It means that the universe passes through the matter-dominated phase. During late time evolution, EoS parameter of HDE becomes
\begin{equation}
w_h=-\frac{1}{3}-\frac{2\sqrt{\Omega_h}}{3 c} -\frac{b^2(1+\Omega_k)}{\Omega_h},
\end{equation}
which gives a negative value. Analysing $w_h$ in (40), one can observe that $w_h$ will definitely cross the phantom divide line in the late time evolution. The late time value of $w_h$ depends on the values of coupling constant $b^2$, $c$, $\Omega_h$ and $\Omega_k$. \\
\indent Now, dividing the Eq. (6) by $H^2$ and substituting the value of $w_h$ from Eq. (40), we get the following value of deceleration parameter for interacting HDE.
\begin{eqnarray}
q & = & \frac{(1+3b^2)(1+\Omega_k)-\Omega_h\left(1+\frac{2\sqrt{\Omega_h}}{c}\cos y\right)}{2\left(1+\frac{\beta a}{(\alpha+\beta a)\; ln(\alpha+\beta a)}\right)}\notag\\
& + & \frac{{\frac{2\beta a(-\Omega_h+3b^2+2)}{(\alpha+\beta a)\; ln(\alpha+\beta a)}}
 -\frac{2\beta^2 a^2}{(\alpha+\beta a)^2\; ln(\alpha+\beta a)}+\frac{2(\omega+1-3\omega b^2)\beta^2 a^2}{(\alpha+\beta a)^2\; [ln(\alpha+\beta a)]^2}}{2\left(1+\frac{\beta a}{(\alpha+\beta a)\; ln(\alpha+\beta a)}\right)},
\end{eqnarray}
\noindent From (41), we observe that the last term will dominate during the evolution due to the large value of the BD parameter $\omega$ provided $1+3b^2>0$, which means that $q$ becomes positive and hence it describes the decelerated phase. In the late-time of the evolution the terms $\frac{\beta a}{(\alpha+\beta a)\; ln(\alpha+\beta a)}$, $\frac{\beta^2 a^2}{(\alpha+\beta a)^2\; ln(\alpha+\beta a)}$ and $\frac{\beta^2 a^2}{(\alpha+\beta a)^2\; [ln(\alpha+\beta a)]^2}$ converge to zero and $\cos y$ converge to $1$,  the value of $q$  is given by
\begin{equation}
q=\frac{(1+3b^2)(1+\Omega_k)-\Omega_h(1+\frac{2 \sqrt{\Omega_h}}{c})}{2}.
\end{equation}
From above we observe that $q$ is negative for $b^2<\frac{\Omega_h(1+\frac{2 \sqrt{\Omega_h}}{c})}{3(1+\Omega_k)}-\frac{1}{3}$.\\
\indent Let us discuss coincidence problem in interaction HDE model. Using (36), (37) and (38) the evolution of $r_1$ can be expressed as
\begin{equation}
\dot r_1=3 H r_1 \left[ w_h+\frac{b^2(1+r_1)^2}{r_1} \right].
\end{equation}
\indent In the above expression the value within the bracket can be positive or negative but for a suitable value of $b^2$, we can get $|w_h+\frac{b^2(1+r_1)^2}{r_1}|<<|w_h|$. Thus, from Eqs.(34) and (43) we can conclude that the energy density ratio $r_1$ may evolve more slowly in interacting HDE model as compare to non-interacting HDE model. This imply that the interaction between DM and HDE plays a vital role to discuss the coincidence problem. As we know that for getting the soft coincidence, the model must satisfy the condition $|\frac{\dot r_1}{r_1}|_0\le H_0$. In our model we can achieve the soft coincidence if $b^2$ satisfy the condition $b^2\le \frac{(1-3w_{h0}) r_{10}}{3(1+r_{10})^2}$. According to the present observational values $r_{10}=3/7$ and $w_{h0}=-1$, we get $b^2\le \frac{7}{25}$. Thus soft coincidence can be achieved at present if $b^2\le \frac{7}{25}$. It can also be concluded that the smaller the value of $b^2$, the energy density ratio may evolve more slowly. This explanation can resolve the problem of cosmic coincidence and it can be checked by taking any suitable small value of $b^2$ along with $w_{h0}=-1$. This represents that the variation in $r_1$ is more slow as compare to the conventional $\Lambda$CDM model. Thus, the coupling constant $b^2$ plays an important role to resolve the cosmic coincidence problem and also this small value of coupling constant is compatible with the observations. This is also analyzed by Feng et al.\cite{feng}. Thus, we observe that interacting HDE along with the logarithmic form of BD scalar field in the framework of BD theory may capable to resolve the cosmic coincidence problem.\\

\section{Conclusion}
\indent In this paper, we have studied non-interacting and interacting HDE model with future event horizon as an infra-red cut-off in the framework of BD theory. Motivated by work \cite{kumar} we have considered the logarithmic form of BD scalar field to discuss the early and late time behaviour of the Universe. This form of BD scalar field gives a time-varying deceleration parameter for non-interacting and interacting HDE models irrespective of the matter content. We have discussed the dynamical view of early and late-time evolution of the Universe with the help of EoS parameter and deceleration parameter. We have also discussed the cosmic coincident problem. The result of both, non-interacting and interacting models are summarize below.\\
\indent In the first case where we have considered the non-interacting HDE model, we have observed that the EoS starts behaving like as its respective form in GR. In late-time of evolution at $c=1$ our model behaves like cosmological constant, at $c>1$ it shows the quintessence region and it mimic like phantom type at $c<1$. Initially the value of DP is negative but during the evolution the very high bound on BD parameter $\omega$ dominates the other terms and may attain the positive value and it is also observed that in late-time it again attain a negative value, i.e., it shows the phase transition from decelerated to accelerated Universe during the evolution. We have also discussed the cosmic coincidence problem. In this by assuming $\omega_h=-2/3$, i.e., $\mid \frac{\dot r_1}{r_1}\mid = 2H_0$, we get the less acute coincidence problem as compare to $\Lambda$CDM.\\
\indent In the second case where we consider the interacting HDE model, we get a time-dependent value of EoS parameter. In early-time and late-time it behaves same as its respective form of GR. In beginning it gives a negative value which is $<-1/3$, this shows the early-time inflation. Due to the presence of the BD parameter $\omega$, which have a very high positive value(according to the observations), in the EoS parameter the value of $\omega_h$ may get some positive value at any time during the evolution depending upon the values of $\alpha$, $\beta$ and $\omega$. This means that during the evolution of the Universe the decelerated phase of the Universe may occur. In the late-time of the evolution it again gives the negative value which is $<-1/3$. This shows the acceleration of the Universe at late-time. Also, in this case we have obtained a time-depending value of the DP. The presence of the BD parameter shows that during the evolution of the Universe under the condition $1+3b^2>0$, $q$ may attain a positive value. In late-time $q$ will show the accelerated Universe under the condition $b^2<\frac{\Omega_h(1+\frac{2\sqrt{\Omega_h}}{c})}{3(1+\Omega_k)}-\frac{1}{3}$. Thus, we conclude that the interacting HDE case in our model shows the phase transition from deceleration to acceleration, which is a good harmony with the current observations. Also in this case we discuss the cosmic coincidence problem. Since the coupling constant plays a vital role to discuss the cosmic coincidence problem and the smaller value of $b^2$ shows that the energy density ratio may evolve more slowly. Here we observe that the soft coincidence can be achieved if $b^2\le \frac{7}{25}$. Thus we can conclude that in the framework of BD theory the interacting HDE with the logarithmic form of BD scalar field may capable to resolve the cosmic coincidence problem.\\


\begin{thebibliography}{9}
\bibitem{riess} A. G. Riess et al., {\it Astron. J.} {\bf 116}, 1009 (1998).
\bibitem{perl} S. Perlmutter, et al., {\it Astrophys. J.} {\bf 517}, 565 (1999).
\bibitem {ben} C. L. Bennett et al.,  {\it Astrophys. J.} {\bf 148}, 1 (2003).
\bibitem{kom} E. Komatsu et al.,  {\it Astrophys. J. Suppl.} {\bf 192}, 18 (2011).
\bibitem{per} W. J. Percival et al., {\it Mon. Not. R. Astro. Soc.} {\bf 401}, 2148 (2010).
\bibitem{car} S. M. Carroll, {\it Living Rev. Relativity} {\bf 4}, 1 (2001).
\bibitem{wein} S. Weinberg, {\it Reviews of Modern Physics} {\bf61}, 1 (1989).
\bibitem{copeland} E.J. Copeland, M. Sami, S. Tsujikawa {\it Int. J. Mod. Phys. D}, {\bf15}, 1753 (2006).
\bibitem{caro} S. M. Carroll, {\it Phys. Rev. Lett.} {\bf81}, 3067 (1998).
\bibitem{chiba} T. Chiba, T. Okabe, M. Yamaguchi, {\it Phys. Rev.D}  {\bf62}, 023511 (2000).
\bibitem{kam} A. Y. Kamenshchik et al., {\it Phys. Lett. B} {\bf 511}, 265 (2001).
\bibitem{hooft} G. 't Hooft, arXiv:gr-qc/9310026 (1993).
\bibitem{suss} L. Susskind, {\it J. Math. Phys.} {\bf 36}, 6377 (1995).
\bibitem{wei} H. Wei and R. G. Cai, {\it Phys. Lett. B} {\bf 660}, 113 (2008).
\bibitem{copoz} S. Copozzielo, {\it Int. J. Mod. Phys. D}  {\bf11}, 483 (2002).
\bibitem{amen} L. Amendola, {\it Phys. Rev. D} {\bf60}, 043501 (1999).
\bibitem{brans} C. Brans and R. H. Dicke, {\it Phys. Rev.} {\bf 124}, 925 (1961).
\bibitem{bert} B. Bertotti, L. Iess and P. Tortora, {\it Nature} {\bf425}, 374 (2003).
\bibitem{li} M. Li, {\it Phys. Lett. B} {\bf 603}, 1 (2004).
\bibitem{hsu} S. D. Hsu, {\it Phys. Lett. B} {\bf594}, 1 (2004).
\bibitem{setare1} M.R. Setare, {\it Phy. Lett. B} {\bf644}, 99 (2007).
\bibitem{setare2} M.R. Setare, {\it Eur. Phys. J. C} {\bf50}, 991 (2007).
\bibitem{ncruz} N. Cruz, S. Lepe, F. Pena, J. Saavedra, {\it Phys. Lett. B} {\bf669}, 271 (2007).
\bibitem{jzhang} J. Zhang, X. Zhang, H. Liu, {\it Phys. Lett. B} {\bf659}, 26 (2008).
\bibitem{sheykhi} A. Sheykhi, {\it Phys. Lett. B} {\bf681}, 205 (2009).
\bibitem{qwu} Q. Wu, Y. G. Gong, A. Z. Wang, and J. S. Alcaniz, {\it Phys. Lett. B} {\bf659}, 34 (2008).
\bibitem{xu} L. Xu, W. Li and J. Lu, {\it Eur. Phys. J. C} {\bf 60}, 135 (2009).
\bibitem{ban} N. Banerjee and D. Pav\'{o}n, {\it Phys. Lett. B} {\bf 647}, 477 (2007).
\bibitem{kumar} P. Kumar and C. P. Singh, {\it Astrophys. Space Sci.} {\bf 362}, 52 (2017).
\bibitem{arik} M. Arik and M. C. Calik, {\it Mod. Phys. Lett. A} {\bf 21}, 1241 (2006).
\bibitem{cali} M. Arik, M. C. Calik and M. B. Sheftel, {\it Int. J. Mod. Phys. D} {\bf 17}, 225 (2008)
\bibitem{M} M. Li, X. D. Li, S. Wang and X. Zhang, {\it J. Cosmol. Astrophys. Phys.} {\bf906},  036 (2009).
\bibitem{huang} Q. G. Huang and M. Li, {\it J. Cosmol. Astrophys. Phys.} {\bf 0408}, 013 (2004).
\bibitem{berto} B. Bertotti, L. Iess and P. Tortora, {\it Nature} {\bf 426}, 374 (2003).
\bibitem{will} C. Will, {\it Living Rev. Relativity} {\bf 9}, 3 (2006).
\bibitem{Xu} L. Xu et al., {\it Mod. Phys. Lett. A} {\bf 25}, 1441 (2010).
\bibitem{aqua} V. Acquaviva et al., {\it Phys. Rev. D} {\bf 71}, 104025 (2005).
\bibitem{wu} F.-Q. Wu and X. Chen, {\it Phys. Rev. D} {\bf 82}, 083003 (2010).
\bibitem{Li} Y.-C. Li, F.-Q. Wu and X. Chen, {\it Phys. Rev. D} {\bf 88}, 084053 (2013).
\bibitem{stein} P. Steinhardt, in {\it Critical Problems in Physics}, edited by V. L. Fitch and D. R. Marlow (Princeton University Press, Princeton, NJ, 1997).
\bibitem{zla} I. Zlatev, L. Wang and P. J. Steinhardt, {\it Phys. Rev Lett.} {\bf 87}, 5 (1999).
\bibitem{karwan} K. Karwan, JCAP {bf05}, 011 (2008).
\bibitem{acq} V. Acquaviva and L. Varde, JCAP {\bf12}, 001 (2007).
\bibitem{feng} C. Feng et al., {\it Phys. Lett B} {\bf 665}, 111 (2008).
\end{thebibliography}
\end{document}